\documentclass[9pt,twocolumn,twoside]{osajnl}
\journal{jocn} 

\setboolean{shortarticle}{false}

\title{Experimental Investigation of Machine Learning based Soft-Failure Management using the Optical Spectrum}

\author[1,*]{Lars E. Kruse}
\author[1]{Sebastian Kühl}
\author[1]{Annika Dochhan}
\author[1]{Stephan Pachnicke}

\affil[1]{Chair of Communications, Kiel University, Kaiserstr. 2, 24143 Kiel, Germany}

\affil[*]{lars.kruse@tf.uni-kiel.de}

\begin{abstract}
The demand for high-speed data is exponentially growing. To conquer this, optical networks underwent significant changes getting more complex and versatile. The increasing complexity necessitates the fault management to be more adaptive to enhance network assurance. In this paper, we experimentally compare the performance of soft-failure management of different machine learning algorithms. We further introduce a machine-learning based soft-failure management framework. It utilizes a variational autoencoder based generative adversarial network (VAE-GAN) running on optical spectral data obtained by optical spectrum analyzers. The framework is able to reliably run on a fraction of available training data as well as identifying unknown failure types. The investigations show, that the VAE-GAN outperforms the other machine learning algorithms when up to 10\% of the total training data is available in identification tasks. Furthermore, the advanced training mechanism for the GAN shows a high F1-score for unknown spectrum identification. The failure localization comparison shows the advantage of a low complexity neural network in combination with a VAE over established machine learning algorithms.
\end{abstract}

\setboolean{displaycopyright}{false} 

\begin{document}

\maketitle

\section{Introduction}
In today's digital era, the demand for high-speed data is experiencing exponential growth. The interconnected nature of our world means that any disruption to optical links not only results in data loss but also leads to service level agreements no longer being met. Consequently, the complexity and dynamism of optical networks are on the rise, necessitating the adoption of automated and dynamic techniques to enhance network assurance. Traditional approaches relying on conservative designs, guaranteed redundancies, and threshold-based fault detection alarms are no longer sufficient. A promising solution lies in leveraging machine learning (ML) algorithms to enable proactive maintenance of future networks \cite{rafique2017cognitive,musumeci2019tutorial}. However, the effectiveness of most machine-learning algorithms relies heavily on a substantial amount of training data for reliable and accurate operation. Optical performance monitoring (OPM) plays a vital role in obtaining such data by facilitating network-wide monitoring, validation, and development of fault management machine learning algorithms. This entails utilizing optical spectrum analyzers (OSAs) at strategic network nodes to extract optical spectrum information for further utilization within machine learning frameworks. Due to the significance of soft failures, i.e. failures that progressively degrade transmission quality and can evolve into hard failures, addressing them has become increasingly crucial. Dealing with such faults can be done in different ways. A connection in which a fault occurs has a reduction in transmission quality. This will show in a reduction in the metric of the quality of transmission (QoT) at the receiver. If this QoT metric is below a certain threshold, a failure is detected. However, with such an approach the identification and localization capabilities are limited. A localization of the fault can be achieved using for example an optical time-domain reflectometry (OTDR) measurement in the field. However, only span-by-span measurements are possible due to the limitations of OTDRs passing optical amplifiers. Furthermore, such an approach comes with high cost for technical staff as well as time consuming measurements. Machine learning has shown to be possible of extending the before-mentioned approaches giving the opportunity to adaptively choose QoT thresholds (fault detection) and recognize patterns of changing QoT metrics (fault identification) \cite{vela2017ber}. \\
\begin{table*}[!h]
	\begin{center}
		\caption{Brief literature comparison for ML-based soft-failure detection (SFD), soft-failure identification (SFI), and soft-failure localization (SFL) based on optical performance monitoring (OPM) data.}
		\label{tab_literature_comparison}
		\begin{tabular}{c c c c c c c c c}
			\hline
			Literature & SFD & SFI & SFL & OPM Data & ML-Algorithm & SFD Acc. & SFI Acc. & SFL Acc. \\
			\hline
			Vela et al. \cite{vela2017ber} &$\surd$ &$\surd$ & & Rx power, BER & Analytical model & 99.06\% & 99.55\% & \\
			Furdek et al. \cite{furdek2021optical} & $\surd$ & &  & BER, block errors, etc. & DBSCAN, SVM & 96.2\% & & \\
			Shariati et al. \cite{shariati2019learning} &$\surd$ &$\surd$ & & Optical spectrum (1 Ch.) & SVM & up to 100\% & up to 100\% & \\
			Lun et al. \cite{lun2020soft} & &$\surd$ & & PSD & CNN & & up to 100\%& \\
			Mayer et al. \cite{mayer2021machine} & & & $\surd$ & Tx Power, OSNR & ANN & & & up to 100\% \\
			Abdelli et al. \cite{abdelli2021reflective} &$\surd$ & & $\surd$ & OTDR & LSTM & 92$\pm$1.06\% & & 2.2$\pm$0.13 m \\
			This paper &$\surd$&$\surd$&$\surd$ & Optical spectrum & VAE-GAN & 99.72\% & 98.21\%&99.82\%\\
		\end{tabular}
	\end{center}
\end{table*}
\indent In recent years, substantial efforts have been devoted by the scientific community to discover more applicable machine learning algorithms for managing soft failures, encompassing soft-failure detection, identification, and localization. In \cite{vela2017ber}, four failures affecting the signal of an optical connection are considered including signal overlap, tight filtering, gradual drift and cyclic shift of filters. The fault detection is achieved using an adaptive threshold mechanism for bit error rate (BER) changes, while the identification is achieved using pattern recognition algorithm. Furdek et al. \cite{furdek2021optical} extend the failure detection capabilities by including more monitoring data in the analysis as well as using an unsupervised learning approach for detecting unknown faults which might include physical-layer attacks. They propose the usage of density-based spatial clustering of applications with noise (DBSCAN) on a dimensional reduced dataset for this purpose. Soft-failures arising from filters are assumed in \cite{shariati2019learning} in which the authors use features extracted from the optical spectrum from one channel to run a support vector machine (SVM) for soft-failure detection and identification. In \cite{lun2020soft}, a one-dimensional convolutional neural network (CNN) is used for soft-failure identification. The CNN is used on the power spectrum density (PSD) extracted from a coherent receiver and trained to identify the cause of variations in the PSD. The authors assume four types of soft-failures, i.e., filter shift, filter tightening, amplified spontaneous emission (ASE) noise increase and increase in Kerr nonlinear effects arising from a launch power increase. In \cite{mayer2021machine}, an artificial neural network (ANN) is used to achieve soft-failure localization in scenarios of partial available telemetry data. The ML algorithm classifies with features based on optical signal-to-noise ratio (OSNR) and transmitter power data and tries to localize amplifier gain degradation, transponder power degradation and additional fiber losses. An ML-assisted OTDR approach is proposed in \cite{abdelli2021reflective}. A long-short term memory (LSTM) network is used to detect a failure in the OTDR backtrace and to localize the fault cause with an mean accuracy of 2.2 meters. A brief summary of corresponding literature is given in Table \ref{tab_literature_comparison}.\\
\indent In most recent work only partial solutions of the entire soft-failure management including soft-failure detection, identification and localization have been provided. This is mainly due to the OPM data used, which is not meaningful enough for all stages, and the ML algorithms used. The ML algorithms used, aside from the ANN trained with partial telemetry data from \cite{mayer2021machine}, all require a large amount of training data, which limits their application in real optical networks.
Furthermore, most of the data are of theoretical or simulative manor, which means no experimental data is underlying those investigations. An exception makes ref. \cite{mayer2022demonstration}, where the authors use theoretical failure data and information to form a digital twin together with a small experimental dataset.\\
\indent In this paper, we extend our work from \cite{kruse2021edfa} and \cite{Kruse2023_ECOC} by comparing our proposed variational autoencoder (VAE) and generative adversarial network (GAN) based framework to other ML algorithms, i.e. a linear classifier, a k-nearest neighbors classifier \cite{cover1967nearest}, a support vector machine based classifier \cite{Smola2004} with a radial bias function kernel, a decision tree classifier \cite{Breiman1984}, a random forest classifier \cite{liaw2002classification}, and a VAE-NN hybrid. On top of that, we set the topic into context and provide more information on the underlying ML algorithms as well as showing the usage of advanced training mechanisms for GAN for identifying unknown failure types. We use experimental emulated soft-failure data to train and evaluate the ML algorithms. The experimental comparison of the ML algorithms shows that soft-failure management can be achieved using the optical spectrum as input data. The VAE-based soft-failure detection using the Euclidean distance as a metric shows to be advantageous over other threshold based detection mechanisms since no exhaustive threshold optimization has to be done. Furthermore, the VAE-GAN approach shows significantly better identification performance on a fraction of the total training data compared to conventional ML solutions. It also shows superior accuracy for unknown failure detection.\\
\indent We show that the benefits of using the optical spectrum as an input for ML algorithms could possibly justify the high deployment costs of OSAs to be used as channel monitors at high priority nodes in an existing network. However, for a greenfield scenario, the deployment of channel monitors at all intermediate nodes might be an option. Furthermore, we showed in \cite{kruse2023experimental} that with the usage of low-resolution (50 pm) OSAs an accurate quality of transmission estimation based on the optical spectrum is possible. Those OSAs come at lower costs and higher update speeds and thus are ideal candidates for deployment at high priority nodes \\
\indent The remainder of this paper is organized as follows: First, a brief overview of the theory of the variational autoencoder and the generative adversarial network with respect to the proposed approach is given in Section \ref{section_background}. Section \ref{section_sfm_framework} contains the description of the proposed soft-failure management framework including the design choices of the ML algorithms. Furthermore, the experimental investigations including the experimental setup and the comparison of the ML algorithms are described in Section \ref{section_experimental_investigations}. A conclusion will be drawn in Section \ref{section_conclusion}.
\section{Background}\label{section_background}
\subsection{Variational Autoencoder}\label{subsection_vae}
\begin{figure}[!t]
	\centering
	\includegraphics[scale=0.63]{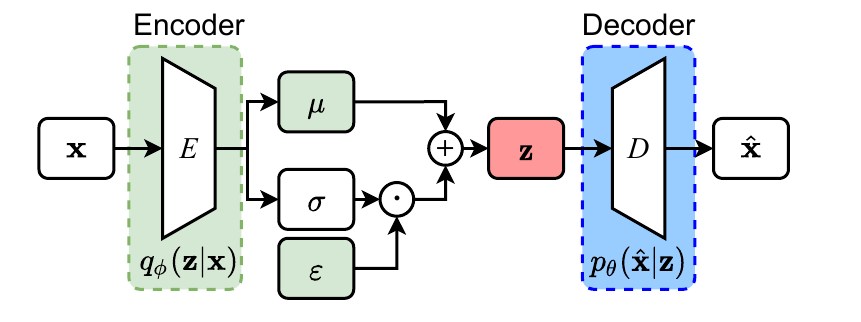}
	\caption{Basic structure of a variational autoencoder (VAE); $\mathbf{x}$: input vector, $\mathbf{z}$: multivariate latent vector, $\hat{\mathbf{x}}$: reconstructed input vector; $\varepsilon$: sample from a normal distribution.}
	\label{fig_vae_structure}
\end{figure}

Autoencoders (AEs) are a type of artificial neural network architecture comprising an encoder $E:X \rightarrow Z$ and a decoder network $D : Z \rightarrow X$, where $ Z \in \mathbb{R}^n$, $n \in \mathbb{N}^+$. These are jointly trained to reconstruct unlabeled data $X \in \mathbb{R}^m$. By selecting a lower dimension $n < m$, represented by the multivariate latent vector $z = E(x)$ with $x \in X$ , the encoder $E$ learns to encode the input data $X$ in a way that enables reconstruction with the decoder $\hat{x} = D(z)$, where $z \in Z$. Trained autoenconders enable applications such as dimensionality reduction by using $z$ instead of $x$, denoising by utilizing $\hat{x}$, and anomaly detection by measuring the discrepancy between $x$ and $\hat{x}$. Kingma and Welling \cite{kingma2019introduction} introduced VAEs as an extension of AEs. VAEs share a similar architecture with AEs, but have a key difference in their objective. Instead of directly reconstructing the data, VAEs aim to learn the distribution of the data using a prior distribution $p_\theta$ parameterized by $\theta$. The latent vector $z$ is typically assumed to follow a multivariate Gaussian distribution. This Gaussian assumption enables additional capabilities beyond conventional AEs, such as data generation by decoding samples drawn from a Gaussian distribution using the probabilistic decoder. Typically, VAEs demonstrate better generalization due to the fact, that encoded samples are not reconstructed directly, but parameterize the distribution from which the input of the decoder is drawn. As the true posterior $p_\theta(z|x)$ is often intractable, it is approximated by a function $q_\phi(z|x) \approx p_\theta(z|x)$ parameterized by
the probabilistic encoder $E_\phi(x)$. The multivariate latent vector is calculated as follows:
\begin{align}
	z = \mu + \sigma \odot \epsilon
\end{align}
where $\mu$ represents the mean value, $\sigma$ is the standard deviation, and $\epsilon$ is a sample drawn from a normal distribution with a mean value of 0 and a standard deviation of 1. The basic structure of a VAE is illustrated in Fig. \ref{fig_vae_structure}. During training, the goal is to find optimal parameters $\theta$ and $\phi$ that minimize the reconstruction error at the decoder output while preserving the Gaussian probability distribution in the latent space. By utilizing a well-trained encoder, the input dimension $m$ can be effectively reduced to $n$ $(n < m)$ with minimal information loss. Thus, the latent space represents a set of meaningful features for describing the input data, which reduces the need for manual feature selection in other machine learning algorithms.
\subsection{Generative Adversarial Network}
\begin{figure}[!t]
	\centering
	\includegraphics[scale=0.9]{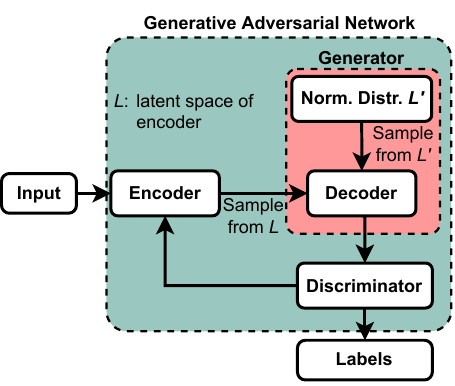}
	\caption{Basic structure of a generative adversarial network (GAN).}
	\label{fig_gan_structure}
\end{figure}

Generative adversarial networks have emerged as a powerful tool for generating realistic and high-quality synthetic data in various domains, including computer vision and natural language processing. GANs, introduced by Goodfellow et al. in 2014 \cite{goodfellow2020generative}, consist of two neural networks: a generator and a discriminator, which compete against each other in zero sum game \cite{goodfellow2020generative}. The basic structure of a GAN is depicted in Fig. \ref{fig_gan_structure}. The generator network aims to produce synthetic data samples that resemble the real data distribution, while the discriminator network strives to differentiate between the real and fake samples. During training, the generator produces synthetic samples, and the discriminator provides feedback by labeling to each sample. The two networks are trained simultaneously in an adversarial manner, with the goal of improving the generator in its ability to deceive the discriminator, and the discriminator being trained to accurately discriminate between real and fake samples. By leveraging this adversarial training process, GANs have been successful in producing images indistinguishable from real photos by humans \cite{creswell2018generative}, synthesizing natural language and generating music. However, GANs are prone to training instability. The hyperparmeters of the networks have to be chosen carefully since GANs suffer from convergence oscillations and vanishing gradients \cite{arjovsky2017wasserstein}. Achieving a balance between the generator and discriminator networks during training can be challenging, leading to suboptimal results or failed convergence.
\begin{figure*}[!t]
	\centering
	\includegraphics[scale=0.75]{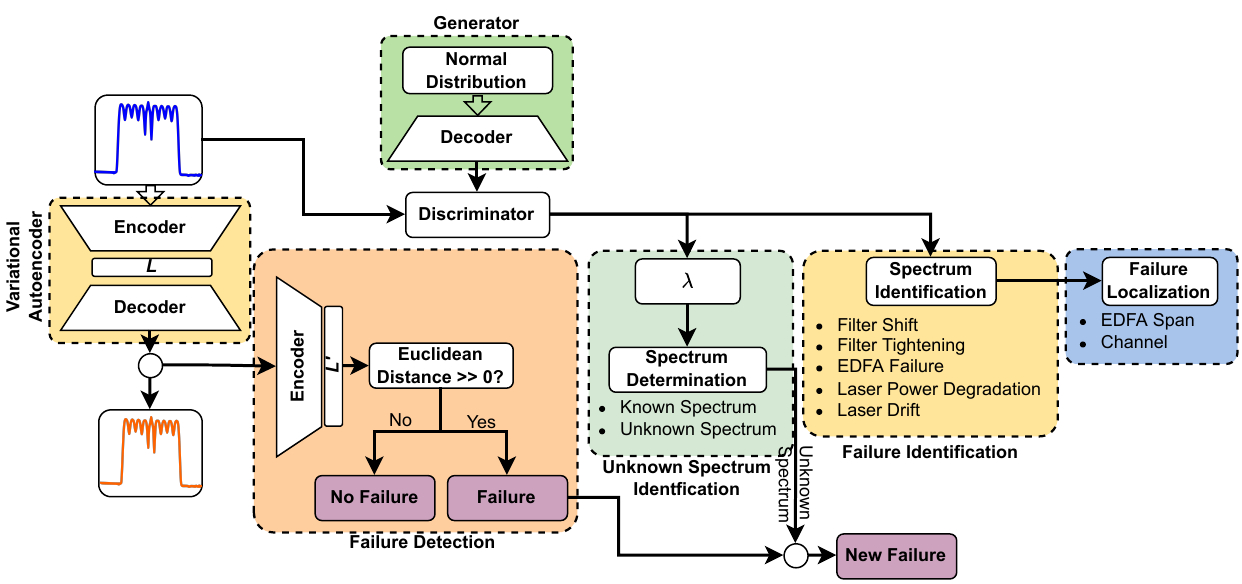}
	\caption{Soft-failure management framework with failure detection, identification, and localization stages in combination with a generative adversarial network (GAN) for unknown spectrum identification; $\lambda$: layer with a custom activation function, $L$: latent space.}
	\label{fig_sfm_framework}
\end{figure*}
\section{Spectral Data driven Soft-Failure Management}\label{section_sfm_framework}
As depicted in Fig. \ref{fig_sfm_framework}, the proposed spectral data driven soft-failure management framework consists of four stages, i.e., failure detection, unknown failure identification, failure identification and failure localization. Previous studies have demonstrated the efficacy of autoencoders in semi-supervised anomaly detection (e.g., \cite{an2015variational,abdelli2022machine}). In this work, we employ a VAE which utilizes stochastic variables as latent variables due to its probabilistic encoder. This feature enhances the VAE's anomaly detection capabilities since normal and anomalous data may exhibit similar mean values but differ in variance meaning another distribution. The stochastic nature of the latent space allows for generating outputs from the decoder by sampling latent space variables from its known normal distribution. Soft-failure detection is accomplished using the aforementioned VAE by calculating the Euclidean distance between the latent space encoded input spectrum (L) and the latent space of the encoded reconstructed spectrum (L'). An anomaly is detected, if the Euclidean distance significantly exceeds zero. This approach is advantageous over threshold-based reconstruction error comparison methods as it eliminates the need for threshold optimization. The VAE's inherent advantages make it suitable for utilization within a GAN to generate more realistic output spectra. In this work, we adopt the approach outlined in \cite{salimans2016improved}, which incorporates both an unsupervised branch for unknown failure identification and a supervised branch for failure identification within the discriminator. Fig. \ref{fig_sfm_framework} also illustrates the two output branches of the discriminator: the lambda layer, i.e. a custom activation function, and the spectrum identification. Initially, a supervised model is established employing a softmax activation function to accommodate five failure classes. Subsequently, an unsupervised model is constructed using the lambda layer that facilitates the implementation of a custom activation function. The lambda layer processes the softmax output from the supervised model and computes a normalized sum of the exponential inputs \cite{salimans2016improved}:\begin{align}
	D(x) = \dfrac{Z(x)}{Z(x)+1},\text{ with }Z(x) = \sum_{k=1}^{K} \text{exp}[l_k (x)],
\end{align} where $D$ is the discriminator, $K$ is the number of classes and $l$ are the logits of the classes. Consequently, the lambda layer's output ranges between 0 and 1, enabling the discrimination between known and unknown samples. Both branches share weights in the hidden layers, thereby establishing a symbiotic relationship where their classification performance is interdependent. During training, the gradient is propagated through the VAE, the generator (utilizing the VAE's decoder), and the discriminator. This training approach ensures that the VAE is optimized not only for superior reconstruction performance but also for separating the latent space in a manner that allows the discriminator to distinguish real samples from unknown ones. For soft-failure localization, a supervised ML algorithm is employed on the input spectrum. The proposed framework underwent extensive optimization using an exhaustive grid search encompassing 80,000 configurations varying the numbers of hidden layers, neurons and using different activation functions. This resulted in the VAE encoder having an input layer of size 501 in order to match the number of spectrum points obtained by the OSA, one hidden layer of size 25, a batch normalization and an output layer of size 12. Due to this, the latent space size is 12 which is also the input layer size for the VAE decoder. The decoder mirrors the encoder. The rectified linear unit (ReLU) function is used as the activation function for the layers. The discriminator of the GAN is composed out of one input layer (size 501), two hidden layers (size 85 and 42, respectively) and an output layer, which is as large as the number of failure classes (i.e. five).
\section{Experimental Investigations}\label{section_experimental_investigations}
While simulations enable the coverage of complex network structures, experiments are essential to validate machine learning algorithms for a possible deployment in a real-world scenario. The data gathered from experimental investigations regarding soft-failures can be used in a potential deployment to approximate the conditions in an optical transmission as well as giving the possibility for future usage of the data for training of machine learning algorithms. Here, we compare the soft-failure detection, identification and localization performances of different ML algorithms to the proposed framework based on GAN with regards to their performance on a fraction of the total training data. As an evaluation metric, the F1-score is chosen, because it provides one concise metric that summarizes the model's performance, even for multi-class scenarios. Furthermore, in situations where the classes are imbalanced (i.e., one class significantly outnumbers the other), which might be the case for soft-failure management tasks, accuracy alone might not be a suitable metric. The F1-score is robust in such cases because it accounts for both false positives and false negatives through incorporating recall and precision, making it suitable for evaluating model performance in imbalanced classification tasks.
\subsection{Experimental Setup}
\begin{figure*}[!h]
	\centering
	\includegraphics[scale=0.7]{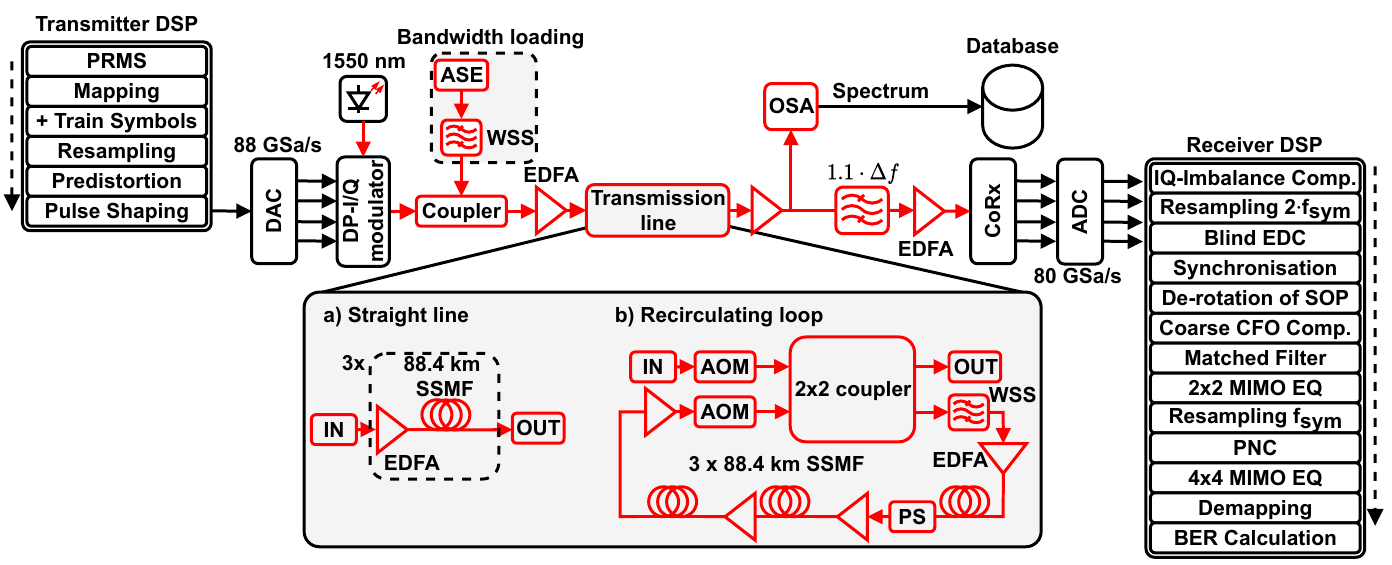}
	\caption{Experimental transmission setup using a) a straight line or b) a recirculating loop. PRMS: pseudo-random multilevel sequence, DAC: digital-to-analog converter, ASE: amplified-spontaneous emission, WSS: wavelength selective switch, EDFA: Erbium-doped fiber amplifier, PS: polarization scrambler, EDC: electrical dispersion compensation, SOP: state of polarization, CFO: carrier frequency offset, PNC: phase-noise compensation.}
	\label{fig_aufbau_labor}
\end{figure*}
The experimental setup, depicted as a high-level black-box model in Fig. \ref{fig_aufbau_labor}, serves the purpose of generating experimental data \cite{kruse2023experimental}. Offline execution of the digital signal processing (DSP) is accomplished through MATLAB routines. To create the channel of interest (COI) at the transmitter side, a pseudo-random multilevel sequence (PRMS) with a length of $2^{17}-1$ is generated. This sequence is then mapped to QPSK, 8-QAM, or 16-QAM symbols, followed by the addition of training symbols for equalization and synchronization. Prior to up-sampling from the symbol rate of 32 GBd to the digital-to-analog converter (DAC) sampling rate (88 GSa/s), the signal undergoes predistortion to account for the characteristics of the electrical amplifier and DAC. Subsequently, the signal is shaped using a root-raised cosine filter with a roll-off factor of 0.2, resulting in an almost rectangular spectrum. Digital-to-analog conversion is achieved by an arbitrary waveform generator (AWG) operating at 88 GSa/s, with an effective number of bits (ENOB) of 5.5 bits. The COI is generated by an external laser at a wavelength of 1550.004 nm, coupled with a DP-IQ modulator driven by the DAC via four driver amplifiers. For the generation of the other wavelength division multiplexing (WDM) channels (loaders), a programmable wavelength-shaping filter (II-VI WS4000A) is employed, utilizing an amplified spontaneous emission (ASE) noise source as input. This process results in shaped ASE noise that represents the WDM channels proximate to the COI. The wavelength-shaping filter has a periodically repeating filter bandwidth aligned with the channel spacing and is configured to equalize all channels at the output. Opting for noise-loading offers advantages over traditional channel generation as it reduces complexity on the transmitter side, requiring only one modulator, one laser, and one DAC. Comparatively, the characteristics of a noise-loaded signal closely resemble those of a conventional WDM signal  \cite{Richter2018}. The COI and loaders are combined using a 3 dB-coupler and subsequently amplified using an Erbium-doped fiber amplifier (EDFA). \\
\indent For generating non-faulty data, the EDFA output is fed into the recirculating loop. The loop is composed of another waveshaper (Finisar WS4000S) being used as a gain-flattening filter followed by three spans and a polarization scrambler (Fig. \ref{fig_aufbau_labor}). Each span consists of an EDFA running at a constant output power of 10.5 dBm, a VOA after the EDFA to get the desired launch power for the following 88.4 km standard single mode fiber (SSMF). After the first span, the polarization scrambler is localized to randomize the polarization shift effects from the fibers. In total, a dataset with 21,600 spectra is obtained for the non-faulty case.\\
\begin{table}[!b]
	\begin{center}
		\caption{Experimentally emulated soft-failures}
		\label{tab_soft_failures}
		\begin{tabular}{c c c}
			\hline
			Soft-failure & Range & Steps \\
			\hline
			EDFA noise figure increase & 0.2 to 2 dB & 0.2 dB\\
			Transmit laser drift & -2.5 to 2.5 GHz & 0.5 GHz\\
			Transmit laser power drop & -2.5 to 2.5 dBm & 0.5 dBm\\
			Filter tightening & 1 to 5 GHz & 1 GHz\\
			Filter shift & -2 to 2 GHz & 1 GHz\\
		\end{tabular}
	\end{center}
\end{table}
\indent Since it is not feasible to physically damage laboratory equipment to simulate faults, the soft-failures are generated by manipulating component controls and utilizing additional components. Thus, five different soft-failure types, i.e., EDFA noise figure (NF) increase, transmit laser shift, transmit laser power drop, filter tightening, and filter shift, are assumed to be possibly present in the link. For the generation of this fault data, a straight line experiment is done using three spans of 88.4 km SSMF in which different failure cases are emulated. The considered soft-failure cases are summarized in Table \ref{tab_soft_failures}. A variable optical attenuator (VOA) is placed at the midstage access of the inline EDFAs to emulated an increase in EDFA noise due to a pump laser degradation. The attenuation of the VOA is varied from 0.2 to 2 dB in 0.2 dB steps. The transmit laser for the center channel is varied from its center frequency by -2.5 to 2.5 GHz in 0.5 GHz steps to emulate a laser drift. To emulate a power drop of a laser, the laser power is decreased by -2.5 to 2.5 dBm in steps of 0.5 dBm. The same procedure is done for different randomly selected channels in the waveshaper which performs the noise shaping of the loaders. For filter tightening, the waveshaper generating the loaders is used to narrow the  channels by 1 to 5 GHz in 1 GHz steps. By shifting the center frequency of the waveshaper from -2 to 2 GHz in 1 GHz steps, filter shift is achieved. Sweeping through the emulated soft-failure parameters from Table \ref{tab_soft_failures} and the experimental parameters from Table \ref{tab_exp_parameters} results in approximately 800 spectra per failure type and a total failure dataset of 5,600 spectra.\\
\begin{table}[!t]
	\begin{center}
		\caption{Experimental system parameters}
		\label{tab_exp_parameters}
		\begin{tabular}{c  c}
			\hline
			Parameter & Value \\
			\hline
			Modulation format &\begin{tabular}{c}DP-QPSK, \\ DP-8-QAM, \\DP-16-QAM \end{tabular}\\
			Symbol rate & 32 Gbaud \\
			Channel spacing & 37.5 GHz\\
			Number of channels & 1, 3, 5\\
			Launch power & -3, -2, -1, 0 dBm\\
			Center wavelength & 1550.004 nm\\
			Loop length & 265.2 km\\
			Loop iterations & 1 to 6\\
			OSA Resolution & 10 pm\\
			OSA Points & 501\\
		\end{tabular}
	\end{center}
\end{table}
\indent At the receiver side, the signal undergoes amplification through an additional EDFA before the COI is filtered. Subsequently, the COI is detected using a coherent receiver. Analog-to-digital conversion (ADC) is achieved using an oscilloscope operating at 80 GSa/s. The received signal is subject to various impairments, which can be categorized as either uncompensated (primarily noise and nonlinearities) or compensated. Compensated disturbances include IQ-skews and IQ-imbalances originating from both the transmitter and receiver, laser phase noise from the transmitter and receiver, chromatic dispersion, polarization mode dispersion (PMD), rotation of the state of polarization (SOP), carrier frequency offset, and laser phase noise from the receiver. Offline receiver DSP is performed using standard algorithms tailored for coherent dual-polarization (DP)-WDM systems \cite{Kruse2023_JLT,Ohlendorf2019}. At the end of the DSP chain, the BER is computed. To obtain the spectrum, an OSA with an optical resolution of 10 pm, such as the Adavantest Q8384, is utilized. The experimental parameters for the investigations are summarized in Table \ref{tab_exp_parameters}.
\subsection{Experimental Results}
\begin{table}[!b]
	\begin{center}
		\caption{VAE-GAN framework performance.}
		\label{tab_framework_performance}
		\begin{tabular}{c c}
			\hline
			Framework stage & Max. F1-score\\
			\hline
			Soft-Failure Detection & 0.9941 \\
			Soft-Failure Identification & 0.9821\\
			Unknown Spectrum Identification & 0.9912\\
		\end{tabular}
	\end{center}
\end{table}
\begin{figure}[!t]
	\centering
	\includegraphics[scale=0.8]{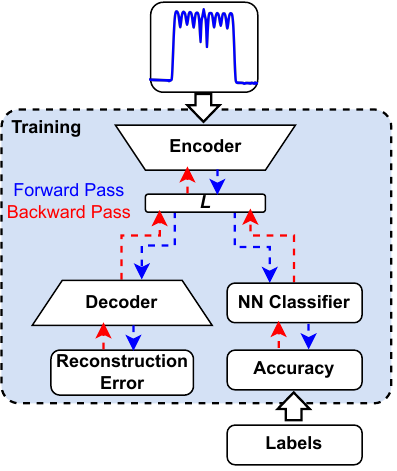}
	\caption{Variational autoencoder and neural network classifier in a two-step training approach.}
	\label{fig_vae_nn_extension}
\end{figure}
The soft-failure detection by the semi-supervised approach using the Euclidean distance reaches an F1-score of 0.9941. The identification stage of the fault achieves an F1-score of 0.9821 while the unknown spectrum identification reaches an F1-score of 0.9912. The results of the framework can be seen in Table \ref{tab_framework_performance}. \\
\indent For evaluating the performance of the proposed framework, we compare it to conventional ML algorithms and a VAE-NN hybrid. The ML algorithms are a linear classifier, a k-nearest neighbors classifier \cite{cover1967nearest}, a support vector machine based classifier \cite{Smola2004} with a radial bias function kernel, a decision tree \cite{Breiman1984}, and a random forest classifier \cite{liaw2002classification}. The structure of the VAE-NN hybrid is depicted in Fig. \ref{fig_vae_nn_extension}. This structure allows for a two step training approach: First, the VAE is trained to optimize the reconstruction error between the input and the output spectrum. As a second step, the VAE's encoder and the neural network based classifier for the soft-failure identification are trained. A mutual beneficial interplay between the encoder and the NN classifier is achieved with the two step training approach. This is due to the failure types are being separated in the latent space while the non-faulty data is clearly separated from the faulty data. Another advantage of this approach is that the NN can be reduced in size, since its input is the latent space resulting in a lower-complex NN-based classifier as well as a faster training time and better generalization capabilities due to the joint training method.
\subsubsection{Detection performance}
Evaluating the detection capabilities of the framework is established with a division of the dataset into  60\% training, 20\% validation, and 20\% test data. We compare two threshold-based soft-failure detection mechanisms which use the VAE as a basis, i.e. based on the mean squared error (MSE) between the input spectrum and the reconstructed spectrum and the Euclidean distance between the latent space of the encoded input spectrum and the latent space of the encoded reconstructed spectrum. The MSE-based soft-failure detection reaches an F1-score of 0.9881 while the Euclidean distance-based mechanism achieves an F1-score of up to 1. This can be explained as follows: If a failure is occurring the Euclidean distance is getting large, i.e. in the $10^4$ regime, while a reconstructed non-faulty spectrum from the VAE shows a Euclidean distance in the range of $10^2$. This means, if a well-trained VAE is used with this approach, a very high accuracy can be reached. However, to cover a wider variation of failures and also being able to detect even small variations in the optical spectrum, we set the threshold to be around $10^2$ resulting in an F1-score of 0.9941. Here, the deviation from a perfect detection arise from small deviations in the dataset.\\
\indent The proposed framework is also capable of detecting unknown failures with the help of the discriminator from the GAN. A spectrum is determined as an unknown failure, if the Euclidean distance is above the set threshold and the lambda layers softmax output is 1. To emulate an unknown failure, we assume a random misalignment of all WDM loaders' frequencies by up to 0.5 GHz. For comparison with the literature, we set up unknown failure detection with a DBSCAN algorithm \cite{furdek2021optical}. The algorithm is optimized with an exhaustive grid search in a way to expect the five different failure classes, meaning it labels outliers as unknown failures. The unsupervised learning algorithm DBSCAN reaches an F1-score of 0.8352. If we test the unknown failure detection capability of the proposed framework, it shows an F1-score of 0.9912. This is due to the fact that the GAN is generating so-called "fake" samples to train the discriminator which enables a high accuracy of identifying unknown spectra as inputs.
\subsubsection{Identification performance}
\begin{figure}[!t]
	\centering
	\includegraphics[width=3in]{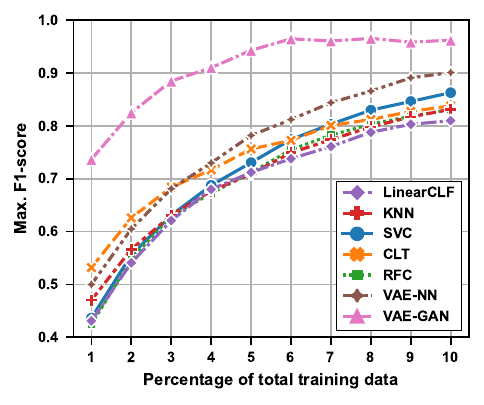}
	\caption{Maximum F1-score for the different machine learning algorithms in the soft-failure identification stage over the percentage of used training data from the total number of training data; LinearCLF: linear classifier, KNN: k-nearest neighbors, SVC: support vector classifier, CLT: decision tree, RFC: random forest classifier.}
	\label{fig_identification_performance}
\end{figure}
\begin{table}[tb]
	\begin{center}
		\caption{Performance of different ML algorithms on total training data.}
		\label{tab_f1_performance}
		\begin{tabular}{c c c}
			\hline
			ML algorithm & \begin{tabular}{c}Identification \\ Max. F1-score \end{tabular}& \begin{tabular}{c}Localization \\ Max. F1-score \end{tabular}\\
			\hline
			Linear CLF & 0.8913 & 0.9872\\
			KNN & 0.9925 & 0.9924\\
			SVC & 0.9943 & 0.9939\\
			CLT & 0.9642 & 0.9723\\
			RFC & 0.9684 & 0.9945\\
			VAE-NN & 0.9973 & 0.9982\\
			VAE-GAN & 0.9821 & /\\
		\end{tabular}
	\end{center}
\end{table}
\begin{table}[tb]
	\begin{center}
		\caption{Execution time of different ML algorithms for soft-failure identification.}
		\label{tab_runtime}
		\begin{tabular}{c c c}
			\hline
			ML algorithm & Training time in s & Prediction time in s\\
			\hline
			Linear CLF & 0.6493 & 0.0035\\
			KNN & 0.1562 & 0.0040 \\
			SVC & 0.6760 & 0.3854 \\
			CLT & 0.6452 & 0.0045 \\
			RFC & 0.7884 & 0.0080 \\
			VAE-NN & 34.3143 & 0.0872\\
			VAE-GAN & 1074.3471 & 0.1285\\
		\end{tabular}
	\end{center}
\end{table}
\indent For an evaluation based on all available training data, the dataset is divided into 60\% training, 20\% validation, and 20\% test data. The ML algorithms are optimized using an exhaustive grid search for an optimal F1-score. The resulting F1-scores are summarized in Table \ref{tab_f1_performance}. It can be seen that the linear classifier reaches the lowest F1-score of 0.8913 followed by the tree structures, i.e., the decision tree with an F1-score of 0.9642 and the random forest classifier reaching an F1-score of 0.9684. The k-nearest neighbors approach achieves an F1-score of 0.9925 which is only overtaken by the support vector machine with radial bias function (RBF) kernel (F1-score: 0.9943) and the VAE-NN hybrid (F1-score: 0.9973). The proposed VAE-GAN framework reaches only an F1-score of 0.9821. The reason for this is that balancing between accuracy of identification and the classification of unknown failures during GAN training needs to be done. This trade-off exists alongside the overall trade-off between the GAN's discriminator and generator.\\
\indent The different ML algorithms run on a desktop computer with an Intel i7-9700K CPU with 32 GB of RAM. The execution times of the algorithms are summarized in Table \ref{tab_runtime}. It can be seen that the prediction time of all algorithms is below one second. The training time of the LinearCLF, KNN, SVC, CLT, and RFC are also below one second, while the VAE-NN hybrid needs over 30 seconds for the training. The VAE-GAN is trained after nearly 18 minutes, however, it has to be noted, that the GAN is trained as long as it needs to reach a given accuracy of the generated "fake" samples.\\
\indent The proposed framework excels when it comes to handling only a fraction of the total training data. For this comparison the algorithms are trained on 1\% to 10\% of the total training data and tested on the rest of the dataset. To show the average expected F1-score of the algorithms for small numbers of training data, 100 sub-datasets are created by drawing random samples from the training data. Each of the (sub-) datasets incorporates the corresponding percentage of the total training data amount and is evaluated individually. These results are depicted in Fig. \ref{fig_identification_performance}. It can be seen, that the overall performance of the conventional ML algorithms in the low percentage range is low. The k-nearest neighbors, random forest classifier and linear classifier show similar performance over the entire range. This shows, that the overall classification task is not simple in a way that linear separability or simple clustering can be done for a high classification accuracy. Furthermore, the decision tree handles a low amount of training data better than most of the other ML algorithms until getting eventually outperformed by the support vector machine at 7\% of the training data. Also, after being able to train on more than 4\% of the training data, the VAE-NN hybrid outperforms all other ML-algorithms reaching an F1-score of 0.902 at 10\%. The proposed VAE-GAN method outperforms all other approaches across all percentages of training data. This is due to the generative nature of the GAN. The interplay of the generator and the discriminator increases the accuracy of the classification, since the discriminator is fed generated spectra by the generator. This approach reaches its maximum at around 6\% of the training data and does not increase, even if all training data is available.
\subsubsection{Localization performance}
\begin{figure}[!t]
	\centering
	\includegraphics[scale=0.7]{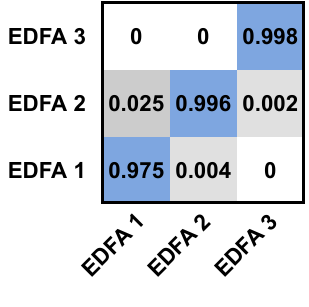}
	\caption{EDFA noise increase localization of SVM with RBF kernel on all training data.}
	\label{fig_edfa_loc_conf_matrix}
\end{figure}
\begin{figure}[!t]
	\centering
	\includegraphics[width=3in]{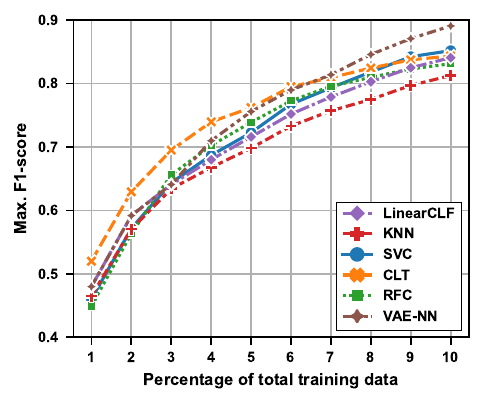}
	\caption{Maximum F1-score for the different machine learning algorithms in the soft-failure localization stage over the percentage of used training data from the total number of training data; LinerCLF: linear classifier, KNN: k-nearest neighbors, SVC: support vector machine based classifier, CLT: decision tree classifier, RFC: random forest classifier.}
	\label{fig_localization_performance}
\end{figure}

For the localization comparison the same ML algorithms are used. However, the VAE-GAN approach is not applicable here, since it is only trained for identifying unknown spectra and identify failures. When considering all training data, the split of the data stays the same as for the identification investigation. The maximum F1-scores are also contained in Table \ref{tab_f1_performance}. The overall F1-scores are high being over 0.97. The decision tree is lowest with an F1-score of 0.9723 followed by the linear classifier. K-nearest neighbor, support vector classifier, random forest classifier, and the VAE-NN hybrid all reaching an F1-score above 0.99 with the VAE-NN hybrid being the best performing algorithm for the localization reaching in an F1-score of 0.9982. The high F1-scores arise from the task itself: Localizing a laser power drop in a channel is straight forward when the optical spectrum is investigated. However, the localization of an EDFA NF increase is more difficult on a per-span level. This can be seen in the exemplary confusion matrix of the support vector classifier in Fig. \ref{fig_edfa_loc_conf_matrix}. Here, the first EDFA is localized with lower accuracy than the last EDFA in the link. This is reasoned by the fact that the SVM with RBF kernel cannot distinguish between an increase in the noise figure of an EDFA or a variation in the noise figure itself. Also, an attenuation of only 0.2 dB in the midstage access of an EDFA results only in a small amount of increased ASE noise. It has to be noted, that for a higher number of EDFAs in the link, the localization accuracy will be lower for the first EDFAs in the link. This may lead to a mislocalization of the underlying issued span. This problem can be addressed by dividing the link into segments, where each segment can contain multiple EDFAs. This approach allows reducing the accuracy error in long links and to increase the precision of localization.\\
\indent If we again assume less training data being available, the algorithms show differences in performance similar to the identification case. As depicted in Fig. \ref{fig_localization_performance}, the k-nearest neighbor algorithm has the lowest overall performance followed by the linear classifier and the random forest classifier. The decision tree again outperforms the other algorithms in the low percentage regime being only outperformed by the VAE-NN hybrid at over 7\% of the total training data which reaches an F1-score of 0.891 at 10\%. The support vector machine classifier also outperforms the CLT at over 9\% of the total training data. Overall all algorithms show a steep trend towards higher F1-scores with more training data being available which mirrors the results before.
\section{Conclusion}\label{section_conclusion}
In this paper, we compared the performance of different ML-algorithms for soft-failure identification and localization using the optical spectrum as input data obtained by sparsely deployed OSAs as channel monitors. Furthermore, we investigated the influence of a lack of available training data on the classification accuracies. The considered ML-algorithms include a linear classifier, a k-nearest neighbors classifier, a support vector machine based classifier with a radial bias function kernel, a decision tree classifier, a random forest classifier, and a VAE-NN hybrid as well as the proposed VAE-GAN framework. The different approaches are compared on experimental soft-failure data acquired for 32 GBaud DP-QPSK, DP-8-QAM, and DP-16-QAM with up to 5 channels with 37.5 GHz spacing over three spans of SSMF. The emulated soft-failures include EDFA noise figure increase, transmit laser frequency drift, transmit laser power drop, filter tightening, and filter shift. The results show, that soft-failure detection, identification and localization as well as unknown spectrum identification is possible on the optical spectrum. Furthermore, the VAE-GAN structure outperforms the conventional ML algorithms when only a fraction training data is available with reaching an soft-failure identification F1-score of 0.9821. For localization purposes, the proposed two-step training approach of the VAE-NN hybrid shows the best performance due to the mutual beneficial interplay between the autoencoder and the neural network. We show that leveraging the generative capabilities of GAN in combination with a VAE enables reliable soft-failure management based on the optical spectrum even with low amounts of training data.\\
\section*{Funding}
CELTIC-NEXT project AI-NET-PROTECT (Project ID C2019/3-4); Bundesministerium für Bildung und Forschung 16KIS1284.

\section*{Acknowledgments}
This work has been performed in the framework of the CELTIC-NEXT project AI-NET-PROTECT, and it is partly funded by the German Federal Ministry of Education and Research.
\section*{Disclosures}
The authors declare no conflicts of interest.

\bibliography{new_journal_bib}



 \section*{Author Biographies}
%
%
%
%
\noindent \textbf{Lars Eike Kruse} received the M.Sc. degree in Electrical and Information Engineering from Kiel University, Kiel, Germany, in 2020, where he is currently pursuing the Ph.D. degree with the Faculty of Engineering. Since 2020, he has been a Research and Teaching Assistant with the Chair of Communications, Kiel University, Kiel, Germany. His current research interests include predictive maintenance and optimization of optical networks with the help of machine learning as well as machine learning based equalization.

\medskip

\noindent \textbf{Sebastian Kühl} received the M.Sc. degree in Computer Science from Bremen University, Bremen, Germany in 2020, and is currently pursuing the Ph.D. degree at the Chair of Communications, Kiel University, Kiel, Germany. The primary focus of his work is the development and evaluation of photonic reservoir computing architectures for signal equalization in fiber optical transmission systems. Past research was centered on the utilization of reinforcement learning for capacity optimization in elastic optical networks.

\medskip

\noindent \textbf{Annika Dochhan} (M'13) received the Dipl. Ing. and Dr. Ing. degrees in Electrical Engineering and Information Technology from Kiel University, Kiel, Germany, in 2004 and 2013, respectively. From 2005 to 2012, she was a research assistant at the Chair of Communications at Kiel University, Kiel, Germany. From 2012 to 2020, she has been with the Advanced Technology Group of ADVA Optical Networking SE, Meiningen, Germany, as a principal engineer responsible for the high-speed transmission lab. From 2020 to 2022 she worked for Dataport AöR, an IT service provider and network operator for public administration in Altenholz, Germany. Since 2022, she is a research staff member with the Chair of Communications, Kiel University, Kiel, Germany, where she is responsible for the optical communications lab. She is involved in many numerical and experimental research activities covering data center networks and WDM systems.

\medskip

\noindent \textbf{Stephan Pachnicke} (M’09–SM’12) received the M.Sc. degree in Information Engineering from City University, London, UK, in 2001, the Dr.-Ing. degree in Electrical Engineering from TU Dortmund, Dortmund, Germany, in 2005, and the Dipl.-Wirt.-Ing. degree in Business Administration from Fern-Universität, Hagen, Germany, in 2005. From 2011 to 2015, he was with ADVA Optical Networking SE in the Advanced Technology Group (CTO Office). Since 2016, he has been a Full Professor and heading the Chair of Communications, Kiel University, Kiel, Germany. He is author or co-author of more than 240 scientific publications, author of a book on Fiber-Optic Transmission Networks (Springer, 2011), and holds several patents. Prof. Pachnicke is
currently serving on the technical program committees of the European Conference on Optical Communications (ECOC), the Optical Fiber Communication Conference (OFC), the Asia Communications and Photonics Conference (ACP) and the International Conference on Photonics in Switching and Computing (PSC). He has been appointed General Chair of the the Signal Processing in Optical Communications Conference (SPPCom) in 2024. He is a member of the European Academy of Sciences and Arts.

%
%
%
%
  
\end{document}